\documentclass[a4paper]{cas-sc}

\usepackage[numbers]{natbib}
\usepackage{threeparttable} 
\usepackage{tabularx}
\usepackage[caption=false,font=normalsize,labelfont=sf,textfont=sf]{subfig}
\newcommand{\etal}{\textit{et al.}\xspace}
\usepackage{graphicx}
\usepackage{url}
\def\tsc#1{\csdef{#1}{\textsc{\lowercase{#1}}\xspace}}
\tsc{WGM}
\tsc{QE}
\tsc{EP}
\tsc{PMS}
\tsc{BEC}
\tsc{DE}

\begin{document}
\let\WriteBookmarks\relax
\def\floatpagepagefraction{1}
\def\textpagefraction{.001}
\shorttitle{PMANet: Malicious URL Detection via Post-Trained Language Model Guided Multi-Level Feature Attention Network}
\shortauthors{Ruitong Liu et~al.}

\title [mode = title]{PMANet: Malicious URL Detection via Post-Trained Language Model Guided Multi-Level Feature Attention Network}                      

\tnotetext[1]{This work was supported in part by the National Natural Science Foundation of China under Grant 61902342, and in part by the Defense Industrial Technology Development Program under Grant JCKY 2021602B002.}


\author[1,2]{Ruitong Liu}
\cormark[0]
\ead{liuruitong@bupt.edu.cn}

\credit{Conceptualization of this study, Methodology, Software}

\affiliation[1]{organization={Key Laboratory of Blockchain and Cyberspace Governance of Zhejiang Province},
                city={Hangzhou},
                postcode={330000}, 
                country={China}}

\author[1,2]{Yanbin Wang }[type=editor,auid=000,bioid=1,orcid=0000-0003-1682-5712]
\cormark[1]
\ead{11921107@zju.edu.cn}

\credit{Data curation, Writing - Original draft preparation}

\affiliation[2]{organization={Zhejiang University},
                postcode={310027}, 
                city={Hangzhou},
                country={China}}

\author[1,2]{Haitao Xu }
\cormark[1]
\ead{ haitaoxu@zju.edu.cn}

\author[1,2]{Zhan Qin }
\cormark[0]
\ead{qinzhan@zju.edu.cn}

\author[1,2]{Fan Zhang }
\cormark[0]
\ead{ fanzhang@zju.edu.cn}

\author[3]{Yiwei Liu }
\cormark[0]
\ead{yiweiliu_disecc@163.com}

\author[4]{Zheng Cao }
\cormark[0]
\ead{z.cao@zju.edu.cn}


\affiliation[3]{organization={Defence Industry Secrecy Examination and Certification Center},
                city={Beijing},
                postcode={065007}, 
                country={China}}

\affiliation[4]{organization={GoPlus Security},
                city={Beijing},
                postcode={065007}, 
                country={China}}
                
\cortext[cor1]{Corresponding author}


\begin{abstract}
The expansion of the Internet has led to the widespread proliferation of malicious URLs, becoming a primary vector for cyber threats. Detecting malicious URLs is now essential for improving network security. The technological revolution spurred by pre-trained language models holds great promise for advancing the detection of malicious URLs. However, current research applying these models to URLs fails to address several crucial factors, including the lack of domain-specific adaptability, the omission of character-level information, and the neglect of both local detail extraction and low-order encoding information. In this paper, we propose PMANet, a pre-trained Language Model-Guided multi-level feature attention network, for addressing these issues. To facilitate a smooth transition of the pre-trained Transformer into the URL domain and to enable it to effectively capture information at both subword and character levels, we propose a post-training program that continues training the model on URLs using three self-supervised learning objectives: masked language model, noisy language model, and domain discrimination task. Subsequently, we develop a module to capture the output of each encoding layer, thus extracting hierarchical representations of URLs spanning from low-level to high-level. In addition, we propose a layer-wise attention mechanism that dynamically assigns weight coefficients to these feature layers based on their relevance. Finally, we apply spatial pyramid pooling to perform multi-scale down-sampling in order to obtain both local features and global context. PMANet achieves multifaceted integration in URL feature extraction, including capturing information at both the lexical and character levels, extracting features from low to high order, and discerning patterns at both global and local scales. We evaluate PMANet against challenging real-world scenarios, such as small-scale data, class imbalance, cross-dataset, adversarial attacks, and case studies on active malicious URLs. All experiments demonstrate that PMANet exhibits superiority over both the previous state-of-the-art pre-trained models and conventional deep learning models. {Specifically, PMANet still achieves a 0.9941 AUC under adversarial attacks and correctly identifies all 20 actively malicious URLs in the case study.} The code and data for our research are available at: \url{https://github.com/Alixyvtte/Malicious-URL-Detection-PMANet}.

\end{abstract}

\begin{keywords}
Malicious URL Detection  \sep Feature Fusion  \sep Transformer \sep Unsupervised Post-training  \sep  Pre-trained Language Model
\end{keywords}

\maketitle

\section{Introduction}

Malicious URLs are deceptive web links designed to facilitate scams and fraudulent activities, with the purpose of convincing users to divulge sensitive and personal information. Clicking on these deceptive links can lead to severe consequences, including personal data theft and cyberattacks. Recent reports, such as Vade's Q1-2023 report \cite{Vade}, indicate a significant increase in phishing attacks, with a quarter-over-quarter volume surge of 102\%, marking the highest Q1 total since 2018. As cyber threats evolve, attackers increasingly spoof trusted brands like Microsoft and Google, rendering malicious URL identification more challenging \cite{cloudflare}. Traditional detection techniques, including blacklist, heuristic, and rule-based systems, suffer from limitations such as delayed updates and inconsistent performance, which are inadequate for the rapid detection of these sophisticated threats \cite{sahoo2017malicious, li2020improving, mamun2016detecting, patgiri2023deepbf}. Consequently, there's a pressing need for more responsive and precise machine learning methods that leverage the distinctive string patterns inherent in malicious URLs for quicker and more accurate threat recognition and classifier training \cite{kim2022phishing, blum2010lexical,sahoo2017malicious}. 

Recent years have witnessed significant advancements in the field of malicious URL detection, largely attributed to the rise of deep learning(DL) techniques\cite{korkmaz2021phishing,maneriker2021urltran,arrieta2020explainable,charte2018practical,peralta2018use,seoni2024application}.  Convolutional neural networks (CNNs) have been extensively employed due to their potent feature extraction capabilities. Notable contributions, such as URLNet\cite{le2018urlnet}, TException\cite{tajaddodianfar2020texception}, and Grambeddings\cite{bozkir2023grambeddings}, have utilized CNNs to deliver exceptional results. However, traditional deep learning approaches still necessitate the manual transformation of URLs into numerical inputs.

Now, the revolutionary pre-trained language model (LM) technologies, such as BERT\cite{devlin2018bert}, have eliminated this last manual step, enabling fully end-to-end learning. Pre-trained LMs are rapidly unifying computational paradigms and frameworks across various domains\cite{islam2023comprehensive}. However, their application to URLs faces several limitations: 1) Domain adaptability issues: While transferring Pre-trained LMs to the URL domain is cost-effective, effectively bridging the gap between general text and the distinct structure of URLs is essential to capture typical patterns of malicious URLs; 2) Lack of character-level information: For malicious URLs, even minor variations or the specific placement of individual characters can be critical indicators of malicious intent; 3) Neglecting multi-orders encoding: Overlooking the granular, hierarchical information encoded in URLs fails to exploit the full spectrum of cues—from broad architecture to nuanced details—that can signal malicious activity; 4) Inability to extract local information: Although Transformers excel at contextual understanding, their performance may be compromised without the capacity to extract and leverage key local features, as threats often depend on subtle local anomalies within a URL.

\begin{figure}
	\centering
	\includegraphics[width=13cm,height=7cm,keepaspectratio=false]{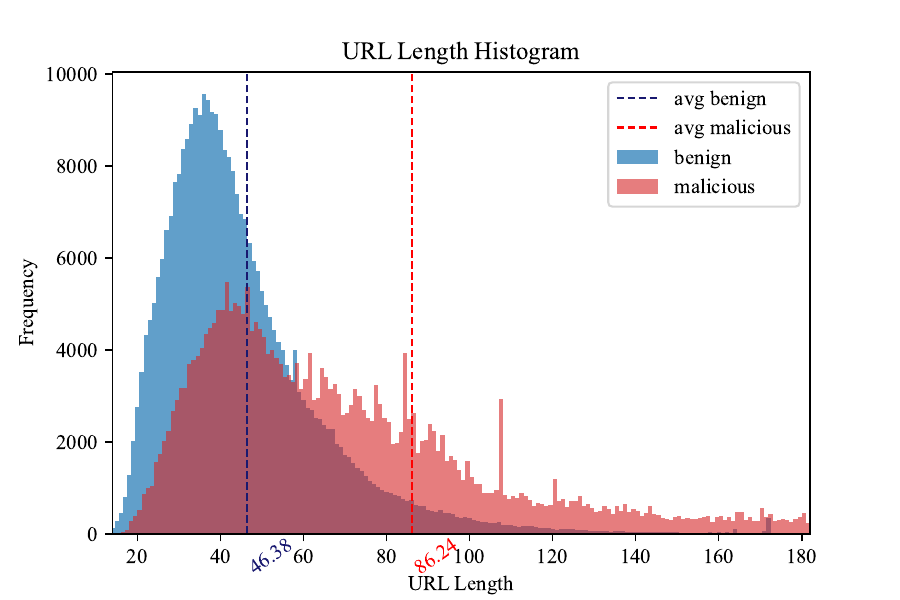}
	\caption{Histogram of URL length.}
    \label{fig:galaxy}
\end{figure}

To address the mentioned challenges, we present PMANet, a pre-trained LM-guided multi-level feature attention network. PMANet utilizes an unsupervised post-training strategy to seamlessly transfer a character-aware pre-trained LM to the URL domain while keeping the learning overhead low. To enrich feature granularity and extract local details, PMANet incorporates a suite of precisely designed modules: a module that dynamically draws embeddings from each layer of the deep encoder, capturing a range of low to high-order information; a layer-aware attention mechanism assigning weights to feature layers according to their importance; and a spatial pyramid pooling module for capturing both intricate local specifics and extensive global context. To rigorously and holistically evaluate our proposed method against the current state-of-the-art technologies, we establish comprehensive and detailed experiments. These not only assess model generalizability and robustness but also evaluate its scalability and responsiveness to data variability. Moreover, we conduct case studies on active malicious URLs to effectively ascertain the real-world performance of our malicious URL detection models.

Our contributions can be summarized as follows:
\begin{itemize}
\setlength{\itemsep}{0pt}

\item Across a series of challenging scenarios, our proposed approach outperforms methods based on previously leading pre-trained models and conventional DL techniques. It achieves superior generalizability and robustness against adversarial attacks while also demonstrating efficient learnability, scalability to varied data volumes, and adaptability to evolving threat landscapes.
\item We propose a post-training program utilizing three unsupervised learning tasks: masked language model, noisy language model, and domain discrimination task—to learn token-level and character-level contexts of URLs, as well as to adapt pre-trained linguistic knowledge to the URL domain.
\item We meticulously design three generic feature processing modules for the pre-trained LM, enabling the pre-trained LM to dynamically weigh and extract multi-level information from URLs, ranging from low-order to high-order, and to capture both local details and global context.

\end{itemize}

In the following, we begin by reviewing the literature in Section \ref{sec:related} and describing the datasets in Section \ref{sec:data}. Detailed explanations of our model's architecture and components are provided in Section \ref{sec:method}. We then present extensive experiments on malicious URL detection and benchmark against baseline methods in Section \ref{sec:experiments}. Finally, a case study is presented and our findings are summarized in Sections \ref{sec:case} and \ref{sec:conclusion}.

\section{Related Work}
\label{sec:related}
Traditional detection methods typically involve the manual extraction of a varied set of elements from URLs for classifier construction \cite{jain2017phishing,almomani2018fast,chiramdasu2021malicious,rupa2021machine,ullah2022malware,ma2011learning,mamun2016detecting}. Given our study's emphasis on data-driven end-to-end learning approaches,  we review two categories of techniques: conventional DL-based methods and those founded on pre-trained LMs.

\begin{table*}
    \centering
    \resizebox{\linewidth}{!}{
    \begin{threeparttable} 
        \captionsetup{position=bottom}
        \caption{The statistics of our dataset for binary classification.}

        \begin{tabular}{lllllllllll}
            \toprule
            \textbf{Dataset}& \multicolumn{3}{l}{\textbf{Sample Sizes}}&  \multicolumn{3}{l}{\textbf{Benign TLDs}}&\multicolumn{3}{l}{\textbf{Malicious TLDs}}\\
            &malicious\tnote{5}  &benign  &total  &.com  &ccTLDs  &other gTLDs    &.com  &ccTLDs  &other gTLDs \\
            \midrule
            \textbf{GramBeddings\tnote{1}}&400,000  &400,000  &800,000  &52.17\%  &12.04\%  &35.79\%  &60.10\%  &11.82\%  &28.08\%   \\
            \textbf{Mendeley\tnote{2}}&35,315  &1,526,619  &1,561,934  &61.97\%  &0.93\%  &37.10\%  &72.86\%  &1.61\%  &25.53\%   \\
            \textbf{Kaggle 1}\tnote{3}&316,251  &316,252  &632,503  &77.46\%  &0.63\%  &21.92\%  &50.59\%  &10.61\%  &38.8\%   \\
            \textbf{Kaggle 2}\tnote{4}&213,037  &428,079  &641,116  &74.27\%  &6.61\%  &19.12\%  &46.62\%  &7.74\%  &45.65\%   \\
            \bottomrule
            \end{tabular}
            
        \begin{tablenotes}[flushleft]
            \item[1,2] \label{fn:binary} These are used for binary classification, download using \href{https://web.cs.hacettepe.edu.tr/~selman/grambeddings-dataset/}{Grambeddings} and \href{https://data.mendeley.com/datasets/gdx3pkwp47/2}{Mendeley} links.
            \item[3] \label{fn:kaggle1} This is used for binary cross dataset test, download using \href{https://www.kaggle.com/datasets/samahsadiq/benign-and-malicious-urls}{this} link.
            \item[4] \label{fn:kaggle2} This is used for multiple classification, download using \href{https://www.kaggle.com/datasets/sid321axn/malicious-urls-dataset}{this} link.
            \item[5] \label{fn:malicious} Indicates malicious URLs in binary test and the total of malicious, defacement, and phishing URLs in multiple test.
        \end{tablenotes}
         \label{tab:dataset}
    \end{threeparttable}}
\end{table*}

\subsection{Conventional DL-based Methods}
Deep learning has significantly advanced malicious URL detection in past years, boosting precision while reducing the need for complex manual feature engineering. In the literature, CNNs are frequently utilized for their robust feature extraction capabilities. A important work in this vein is URLNet \cite{le2018urlnet}, which pioneered the use of dual-path CNNs to learn character and word embeddings—a technique widely embraced by subsequent research. {Other studies \cite{srinivasan2021durld,yan2020learning,wang2022tcurl} have explored hybrid architectures that combine CNNs with other network types, such as RNNs and LSTMs networks. CNNs excel at extracting local features and capturing spatial hierarchies in data, making them particularly effective for identifying patterns in the structure and content of URLs. RNNs and LSTMs, on the other hand, are designed to model sequential data and capture long-range dependencies. This makes them well-suited for understanding the context of different URL components. These hybrid models aim to leverage the strengths of different network architectures to improve overall performance in URL classification tasks.} Bozkir et al. \cite{bozkir2023grambeddings} implemented Grambeddings, which integrate CNNs, LSTMs, and attention mechanisms to capture n-gram-based insights. Huang et al. \cite{huang2019phishing} proposed a four-branch neural network with convolutions and capsule layers for detailed classification, while Wang et al. \cite{wang2019bidirectional} introduced a bidirectional LSTM that fuses CNNs and RNNs with Word2Vec training to efficiently pinpoint static lexical features within URLs, enhancing the detection of malicious URLs\cite{gniewkowski2023sec2vec}. Although these methods advanced the automation of URL representation, they still require manual initialization at the character, word, or n-gram levels. Moreover, the potential of traditional DL seems to have been maximized, with recent efforts struggling to make performance gains.

\subsection{Pre-trained LMs-based Methods}
Pre-trained LMs utilize sophisticated Transformer architectures, learning from countless unlabeled samples and achieving significant progress in various fields. Pre-trained LMs present a fully data-driven methodology, automating the entire data processing pipeline.

Several studies, referenced in \cite{maneriker2021urltran,chang2021research}, have experimented with pre-trained methods for malicious URL detection, adapting BERT models originally trained on English text to recognize malicious URLs. Nevertheless, these direct applications sometimes struggle with domain misalignment due to insufficient feature learning tailored to the specific needs of malicious URL detection. Research \cite{wang2023lightweight} integrated Transformers with a hybrid expert network for exceptional URL classification, but did not fully exploit potential of unsupervised learning. On the other hand, research \cite{wang2023large} constructed a domain-adapted BERT model for URLs from the ground up. Starting anew brings multiple benefits but at the considerable cost of needing vast datasets, considerable computational power, and considerable training periods.

Compared to previous work, our research delves into the key challenges and limitations of applying pre-trained LMs to URLs, identifying critical areas such as domain relevance, contextual understanding, character-level discrepancies, and feature extraction intricacies. Our work bridges the gaps in prior studies, paving the way for the effective utilization of pre-trained LMs in the realm of URL applications by addressing these challenges through innovative model adaptation and training techniques.

\section{Malicious URL Data}
\label{sec:data}

To evaluate our proposed method, we utilize four publicly available datasets, each differing significantly in terms of sample sizes and the distribution of Top-Level Domains (TLDs) among benign and malicious URLs, as illustrated in Figure \ref{fig:galaxy} and Table \ref{tab:dataset}.

\textbf{GramBeddings Dataset:} Provided by Grambeddings \cite{bozkir2023grambeddings}, this dataset consists of 800,000 samples, split evenly with 400,000 malicious and 400,000 benign URLs. Malicious URLs were collected from repositories such as PhishTank and OpenPhish, spanning from May 2019 to June 2021. Benign URLs were acquired through web crawling of well-known websites listed in Alexa, applying down-sampling techniques to maintain balance. Among the benign URLs, 52.17\% used the \textit{.com} TLD, 12.04\% were ccTLDs, and 35.79\% other gTLDs. The malicious URLs comprised 60.10\% \textit{.com}, 11.82\% ccTLDs, and 28.08\% other gTLDs.

\textbf{Mendeley Dataset:} 
Sourced from Mendeley Data\cite{singh2020malicious}, this dataset includes 1,561,934 samples, with a substantial skew towards benign URLs (1,526,619) compared to 35,315 malicious ones. The URLs were collected using the MalCrawler tool and validated through the Google Safe Browsing API \cite{safe-browsing}, presenting a significant class imbalance with a ratio of approximately 1:43. In this dataset, 61.97\% of benign URLs and 72.86\% of malicious URLs used \textit{.com} TLDs, with the rest distributed among ccTLDs and other gTLDs.

\textbf{Kaggle Dataset:} The Kaggle 1 and Kaggle 2 datasets are both obtained from Kaggle. The Kaggle 1 comprises 632,503 samples, equally divided between malicious and benign URLs. This dataset's balanced nature contrasts with the Mendeley dataset, offering a different perspective for evaluation. Here, the benign URLs primarily used \textit{.com} TLDs (77.46\%), with 21.92\% in other gTLDs and a minimal fraction of 0.63\% in ccTLDs. The malicious URLs showed a distribution of 50.59\% under \textit{.com}, 38.8\% under other gTLDs, and 10.61\% in ccTLDs. 

\textbf{Kaggle 2 dataset:}The Kaggle 2 dataset is multi-class, comprising benign positive samples (428,079) and three negative sample categories: defacement (95,306), phishing (94,086), and malicious (23,645). In this dataset, the \textit{.com} TLDs frequency is still high in benign URLs, reaching 74.27\%. The ccTLDs frequency(6.61\%) is slightly higher than in the second and third datasets, with other gTLDs accounting for 19.12\%. In combined malicious URLs across three negative sample types, the \textit{.com} TLDs frequency is 46.62\%, including 7.74\% ccTLDs and 45.65\% other gTLDs, still remaining lower than in benign instances.

The unique composition and TLD distribution of each dataset provided a comprehensive basis for evaluating the effectiveness of our method across diverse web domains, allowing for robust testing under various real-world scenarios.

\section{Method of PMANet}
\label{sec:method}

PMANet leverages a 12-layer encoder architecture of the pre-trained CharBERT as its backbone, integrating three advanced feature processing modules: a multi-order feature extraction module, a layer-aware attention module, and a spatial pyramid pooling module. PMANet employs a two-stage training process: the first stage involves unsupervised continuation of pre-training CharBERT using our innovative post-training program, which utilizes masked language model(MLM) tasks, noisy language model(Noisy ML) tasks, and domain discrimination (DD) tasks; the second stage involves supervised fine-tuning. It's important to note that our post-training program differs from initial pre-training in that the goal is transferring the pre-trained model and its learned knowledge to the target domain. For clarity on PMANet, our description is split into two main subsections: one elaborating on the network structure and another detailing our proposed post-training approach.

\begin{figure}
    \centering
    \includegraphics[width=0.9\textwidth]{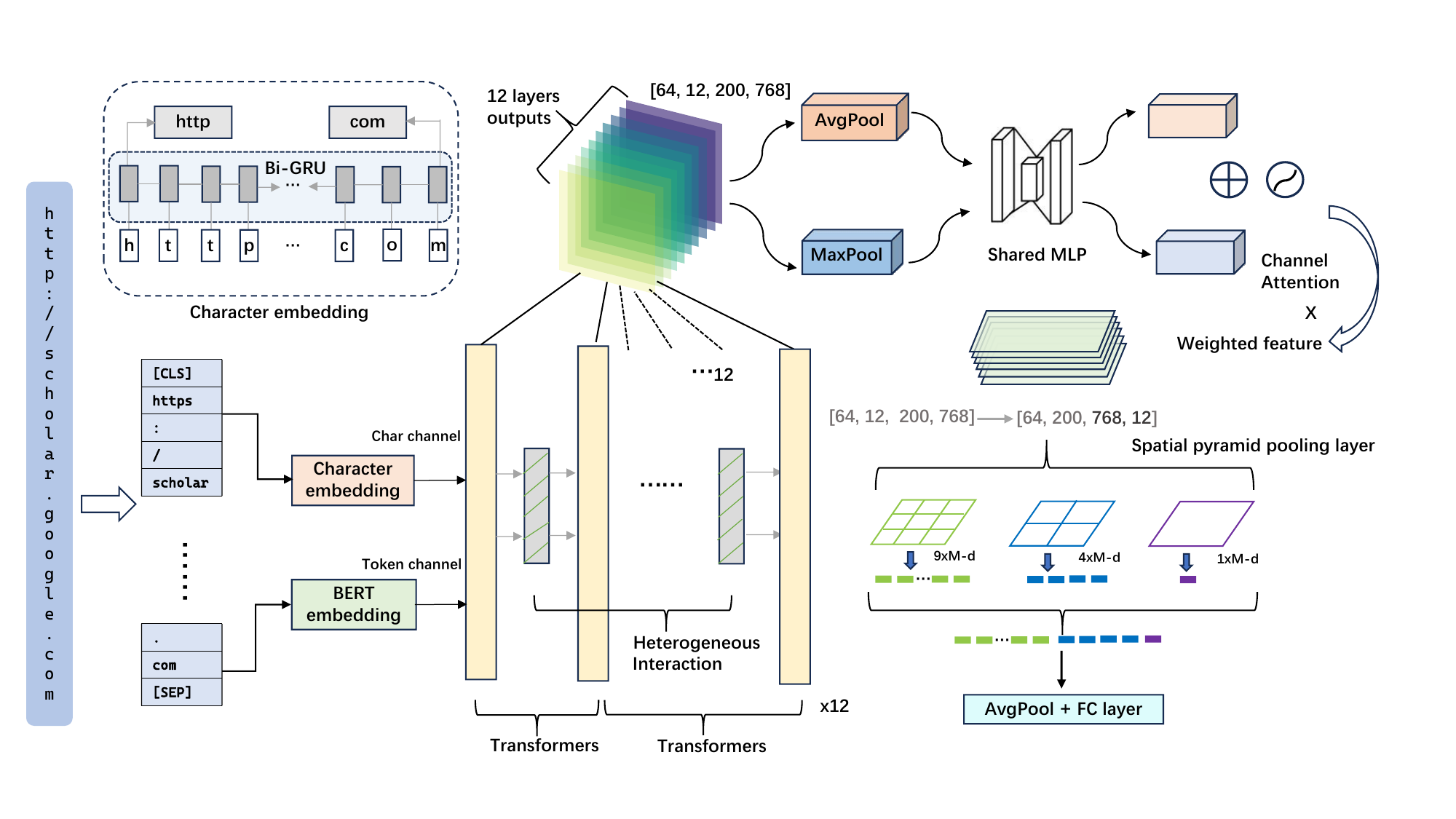}
    \caption{The overall workflow of PMANet. PMANet employs post-trained CharBERT on URLs to extract features at both the character and subword levels. Multi-order feature extraction modules then derive encoded representations ranging from low to high-level. Subsequently, layer-aware attention dynamically readjusts the weighting of different features, which is followed by spatial pyramid pooling that accentuates local nuances and consolidates global context.}
    \label{fig:your_label}
\end{figure}

\subsection{Network Structure of PMANet}
Here, we describe PMANet, including its backbone architecture and the three integrated feature modules that augment the network, as shown in Figure~\ref{fig:your_label}.

\textbf{Backbone Network:} We use the pre-trained CharBERT \cite{ma2020charbert} as our backbone network, which features a dual-channel framework designed to capture information at both the subword and character levels. It consists of two main module: (1) the Character Embedding (CM) Module, which encodes character sequences derived from input tokens, and (2) the Heterogeneous Interaction (HI) Module, which facilitates model to merge and process information effectively from both channels. 

\begin{figure}
    \centering
    \includegraphics[width=0.50\textwidth, ]{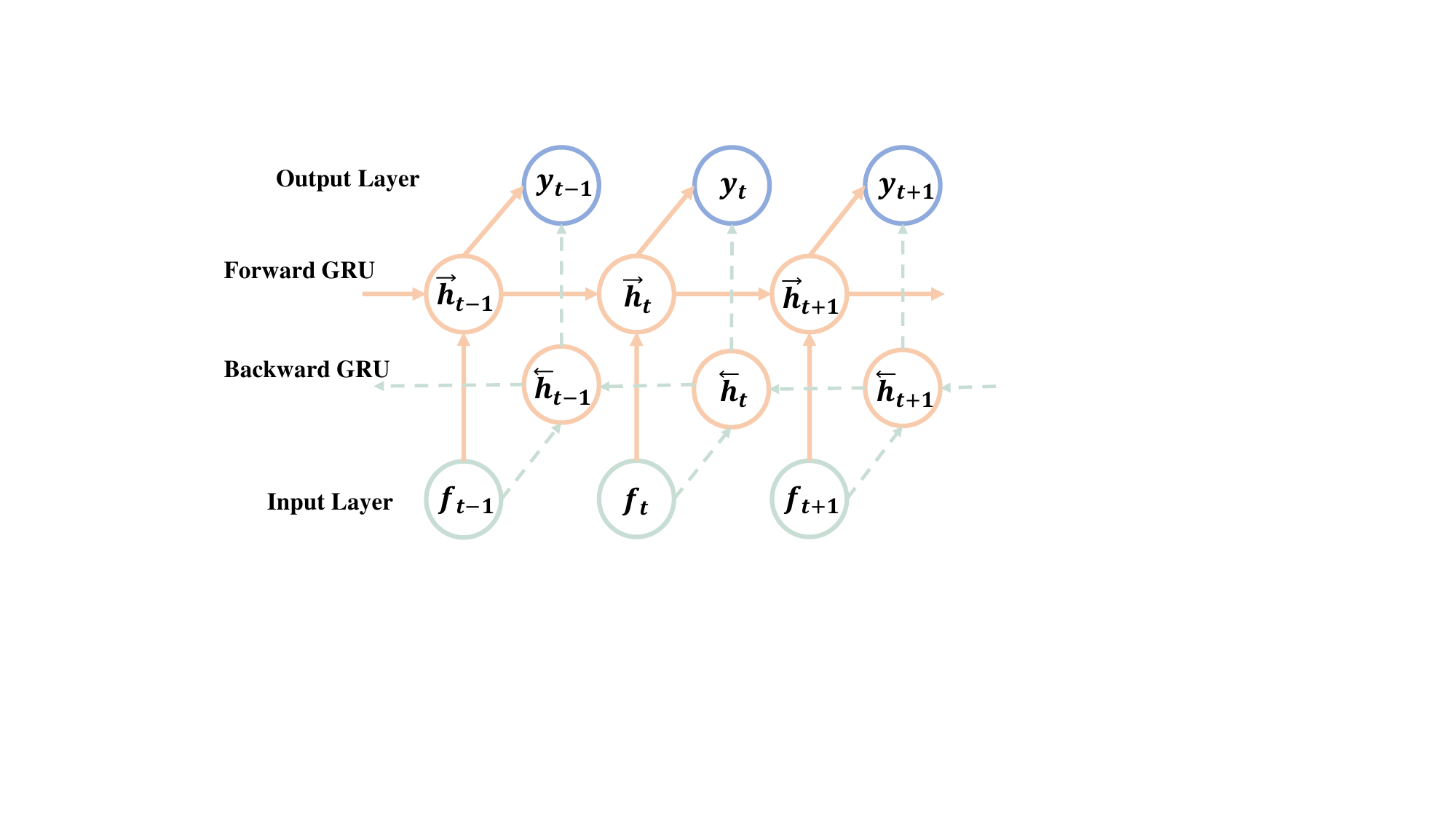}
    \caption{The architecture diagram of BiGRU.}
    \label{fig:architecture}
\end{figure}

The CM module comprises a two-layer bidirectional Gated Recurrent Unit (GRU) network\cite{deng2019sequence}, which integrates hidden layer states from both forward and backward directions to generate context-aware character embeddings. The architecture diagram of the BiGRU is illustrated in Figure \ref{fig:architecture}. {Given an input sequence $f_t (t=1,2,\ldots,d)$ with a window size of $d$, where $f_t$ represents the input feature vector at time step $t$, the forward GRU processes this sequence.}  The resulting forward output sequence of the hidden layer is then depicted by Eq.(1):

\begin{equation}
\overset{\xrightarrow{}}{h_t} = \overset{\xrightarrow{\hspace{0.5cm}}}{\text{GRU}}(\overset{\xrightarrow{\hspace{0.3cm}}}{h_{t-1}},f_t)(t = 1, 2, \dots, d)
\end{equation}
{where $\overset{\xrightarrow{\hspace{0.5cm}}}{GRU}$ represents the mapping relationship of the forward GRU, and $\overset{\xrightarrow{}}{h_t}$ is the forward hidden state at time step $t$.} Similarly, for the input of the backward GRU, the backward output is illustrated in Eq.(2):

\setlength{\abovedisplayskip}{0.1ex}
\setlength{\belowdisplayskip}{0.1ex}
\begin{equation}
\overset{\xleftarrow{}}{h_t}=\overset{\xleftarrow{\hspace{0.5cm}}}{GRU}(\overset{\xleftarrow{\hspace{0.3cm}}}{h_{t+1}},f_t)(t=d,d-1,\dots,1) 
\end{equation}
where $\overset{\xleftarrow{\hspace{0.5cm}}}{GRU}$ is the mapping relationship of the backward GRU, {and $\overset{\xleftarrow{}}{h_t}$ is the backward hidden state at time step $t$.} Then the output of the hidden layer at $t$ can be characterized as:

\begin{equation}
h_t=[\overset{\xrightarrow{}}{h_t},\overset{\xleftarrow{}}{h_t}]=BiGRU(f_t)
\end{equation}
{Here, $h_t$ represents the concatenation of the forward and backward hidden states, denoted by the square brackets $[\cdot,\cdot]$. The $BiGRU$ function represents the operation of the bidirectional GRU, which processes the input $f_t$ in both forward and backward directions. It's worth noting that in Eq.(3), $f_t$ is the same input used in Eq.(1) and Eq.(2), and $t$ ranges from 1 to $d$, covering the entire input sequence.}

To generate character-level embeddings, each input sequence is represented as ${w_1, w_2, ..., w_i, ..., w_m}$, with each $w_i$ being a subword tokenized by BPE and $m$ is the sequence's total subword count. Subwords consist of characters ${c^i_1,...,c^i_{n_i}}$, where $n_i$ is the subword length. We calculate the total length of character input as $N = \sum_{i=1}^{m} n_i$. The processing is outlined as follows:

\begin{equation}
e^{i}_{j}=W_c \cdot c^{i}_{j};\;h^{i}_{j}=BiGRU(e^{i}_{j});
\end{equation}
Where $W_c$ is the character embedding matrix, and $h^{i}_{j}$ represents the representation of the $j$-th character within the $i$-th token. The BiGRU processes characters across the entire input sequence of length N to construct subword-level embeddings. And the hidden states of the first and last characters within each token are then concatenated as follows:

\begin{equation}
h_i(x)=[h^{i}_{1}(x);h^{i}_{n_i}(x)]
\end{equation}
Where $n_i$ represents the length of the $i$-th token, and $h_i(x)$ denotes the subword-level embedding derived from characters. 

The HI module combines and then separates subword and character representations after each Transformer layer. The process involves transforming them separately with dedicated fully-connected layers, before concatenating them using a CNN layer, which is formulated as:
\vspace{10pt}
n\begin{equation}
t^{'}_{i}(x)=W_1*t_i(x)+b ; \; h^{'}_{i}(x)=W_2*h_i(x)+b_2
\end{equation}
\begin{equation}
w_i(x)=[t^{'}_{i}(x);h^{'}_{i}(x)] ; \; m_{j,t}=tanh(W^{j}_{3*w_{t:t+s_j-1}}+b^{j}_3)
\end{equation}
where $t_i(x)$ signifies the token representations, $W$ and $b$ are parameters, $w_{t:t+s_j-1}$ denotes the concatenation of embeddings corresponding to $(w_t,...,w_{t+s_j-1})$, with $s_j$ representing the window size of the $j$th filter, and $m$ is the fused representation, with a dimensionality matching the number of filters.

To separate the fused representations back into two distinct channels, another fully connected layer is employed with GELU activation \cite{hendrycks2016gaussian}, followed by a residual connection to retain specific information from each channel.

\begin{equation}
m^{t}_{i}(x)=\Delta(W_4*m_i(x)+b_4); \; m^{h}_i(x)=\Delta(W_5*m_i(x)+b_5)
\end{equation}
\begin{equation}
T_i(x) = t_i(x)+m^{t}_{i}(x); \; H_i(x)=h_i(x)+m^{h}_{i}(x)
\end{equation}
$\Delta$ is the activation function GELU, T and H are the representations of the two channels. Finally, a layer normalization operation is applied after the residual connection. Fusion enhances mutual representation enrichment, while division preserves distinct token and character features and fosters dual-channel differentiation through pre-training tasks.
\begin{figure}
    \centering
    \includegraphics[width=0.450\textwidth, height=0.250\textheight]{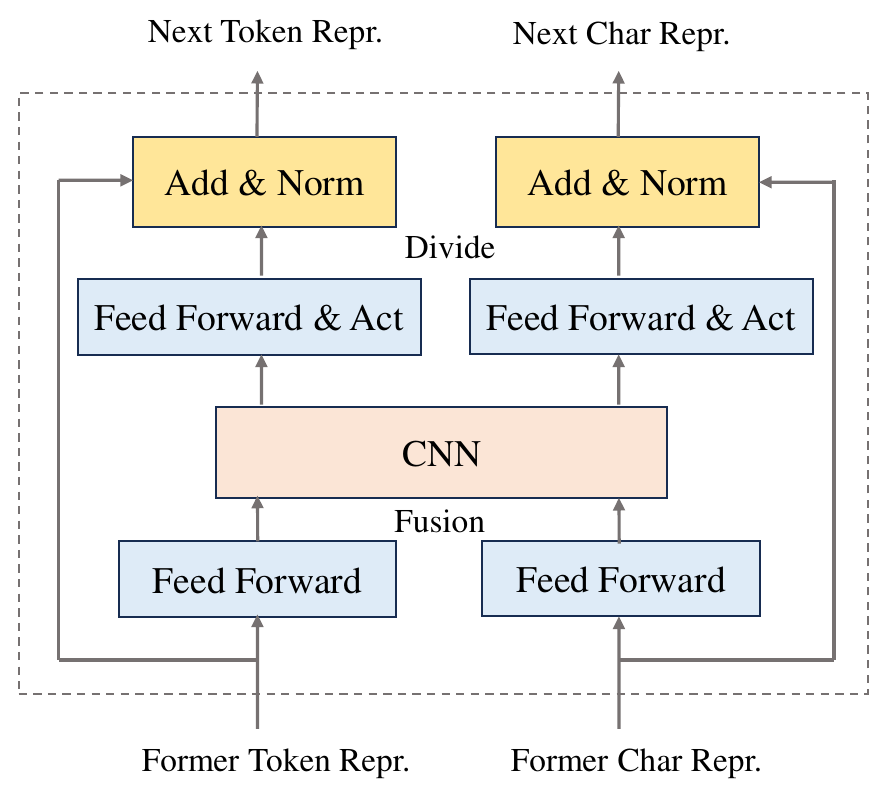}
    \caption{The architecture diagram of Heterogeneous Interaction.}
    \label{}
\end{figure}

\textbf{Multi-order Feature Extraction Module:} Pre-trained LMs like BERT typically stack multiple layers of Transformer encoders, acquiring semantic knowledge through extensive corpus training, and are then fine-tuned for specific downstream tasks. Traditional BERT-based classification models often rely on the final layer's classification feature, denoted as [CLS], encapsulating the semantic information of the entire input sequence. However, Jawahar \etal 's research \cite{jawahar2019does} suggests that BERT can learn extensive information across layers, capturing phrase-level details in lower-level encodings, syntactic information in the middle, and rich semantic features at higher levels. They employed k-means clustering across BERT layer representations and assessed cluster quality using Normalized Mutual Information (NMI). The study found lower BERT layers excel at encoding phrase-level information, evident in higher NMI scores, as shown in Table \ref{tab:jawahar2019}. Deeper BERT layers are more effective in handling long-range dependency information.

Yet, in the context of the BERT model, even as each layer inputs the output features of the preceding one, the complex computations within each layer may lead to degradation of low and middle-level features, hindering the complete feature learning process. This understanding also applies to the context of CharBERT transfer learning. In feature concatenation, Li \etal \cite{Li2020} integrated aspect features from each layer in the context of aspect term sentiment classification using BERT. Compared to relying solely on the final layer's classification feature, this approach significantly enhances classification performance by leveraging different features learned at each BERT layer.

\begin{table}
    \centering
    \caption{Clustering performance of span representations obtained from different layers of CharBERT.}
    \begin{tabular}{cccccccc}
    \toprule
         Layer & 1 & 2 & 3 & 4 & 5 & 6 &  \\
         NMI & 0.38 & 0.37 & 0.35 & 0.3 & 0.24 & 0.2 & \\
    \midrule
         Layer & 7 & 8 & 9 & 10 & 11 &12 \\
        NMI & 0.19 & 0.16 & 0.17 & 0.18 & 0.16 & 0.19 \\
    \bottomrule
    \end{tabular}
    \label{tab:jawahar2019}
\end{table}

The Multi-order Feature Extraction (MoFE) Module is designed to exploit the hierarchical information of CharBERT, extracting granular insights from the low to high-layer representations. The feature extraction algorithm proceeds as follows:

We consider two output sequences: $k_1, k_2, ..., k_n$ and $c_1, c_2, ..., c_m$, where each output $k_i$ and $c_j$ has a rank of $(H, W, C)$, representing the outputs of CharBERT’s word-level and character-level encoders at various layers. We then merge these two channels through a one-dimensional convolution post concatenation, restoring them to their original dimensions. Subsequently, we stack these merged outputs along the new dimension 0 to form a tensor $X$ of rank $(N, H, W, C)$, where $N$ denotes the number of layers (12 in our case), $H$ the batch size, $W$ the fixed URL sequence length (set to 200 in our model), and $C$ a 768-dimensional vector representing the output of each merged hidden layer in CharBERT. To further optimize the representation, a permutation matrix $P$ is defined to rearrange the tensor elements for more efficient processing, ensuring that the multi-level features are optimally aligned for subsequent analytical tasks. We define a permutation matrix $P$ as follows: 

\begin{equation}
  P[i][j] \quad
  \text{
    $\left\{
    \begin{array}{ll}
      1 & \text{if } i=0,j=1;i=1,j=0 \\
      0 & \text{if } otherwise
    \end{array}
    \right.$
  }
\end{equation}

By swapping dimension 0 and dimension 1, we obtain a tensor $X'$ of rank $(H, N, W, C)$. This operation can be represented as $X' = P * X$. Finally, $X'$ serves as the input to the subsequent attention module after the concatenation of the feature layers.

\textbf{Layer-aware Attention Module:}
To highlight the importance of features across various layers, we develop a Layer-aware Attention(LaA), drawing inspiration from channel attention principles. The LaA module empowers the model to independently discern and assign differentiated weights to feature maps at different layers, thus boosting both processing efficiency and precision. This yields two unique spatial context descriptors, $F^{c}{avg}$ and $F^{c}{max}$, for the average pooled and max pooled features, respectively. Subsequently, these descriptors are processed through a shared Multi-Layer Perceptron (MLP) network, which includes a hidden layer, to produce a channel attention map $M \in R^{C\times1\times1}$. Following the application of this shared network to each descriptor, the emerging output feature vectors are merged using an element-wise summation. The methodology is described as follows:

\begin{equation}
\begin{aligned}
M_c(F) &= \sigma(MLP(AvgPool(F) + MLP(MaxPool(F))) \\
&= \sigma(W_1(W_0(F^{c}_{avg})) + W_1(W_0(F^{c}_{max})))
\end{aligned}
\end{equation}
\vspace{10pt}

{Here, $F \in \mathbb{R}$ represents the input feature map, where $C$ is the number of channels (corresponding to the number of layers in the model). The superscript $c$ in $F^{c}{avg}$ and $F^{c}{max}$ indicates that these are channel-wise descriptors. $\sigma$ denotes the sigmoid function.} 

In our experiments, the channel number $C$ is set to 12 as determined in the preceding process, and $r$ is configured to 3.{The matrices $W_0 \in \mathbb{R}^{(C/r) \times C}$ and $W_1 \in \mathbb{R}^{C \times (C/r)}$ represent the weights of the MLP and are shared for both inputs.} Following $W_0$, we apply the ReLU activation function. Subsequently, the attention maps are multiplied with the input pyramid feature map.

\begin{table}

    \footnotesize
    \caption{Impact of number of layers on model performance in terms of accuracy scores.}
    \setlength{\tabcolsep}{13pt}
    \begin{tabular}{cccccc}
    \toprule
         \ layers(count)& Accuracy &Precision  &Recall  &F1-score  &AUC \\
    \midrule
         2&  0.9772&  0.9811&  0.9730&  0.9771& 0.9959 \\
         3&  0.9837&  0.9873&  0.9799&  0.9863&  0.9963\\
         4&  0.9856&  0.9917&  0.9792&  0.9854&  0.9938\\
         5&  0.9860&  0.9861&  0.9858&  0.9859&  0.9998\\
         \textbf{12}&  \textbf{0.9915}&  \textbf{0.9949}&  \textbf{0.9880}&  \textbf{0.9914}&  \textbf{0.9965}\\
    \bottomrule
    \end{tabular}
    \label{tab:tab3}
\end{table}

\textbf{Spatial Pyramid Pooling:}
We apply Spatial Pyramid Pooling (SPP) to the weighted feature results. Originally utilized in computer vision tasks and convolutional neural networks, SPP segments feature maps into locally spatial partitions from fine to coarse levels, aggregating local features and thus becoming a key component in classification and detection systems. We innovatively combine SPP with Transformer technology, applying it to the weighted features extracted by our Layer-Aware Attention module. Specifically, for feature maps of size $a \times b$, we employ $n \times n$ pyramid-level pooling, using a sliding window of size $win = [a/n]$ and stride $str = [b/n]$. Our model incorporates a three-level pyramid.

In the final stage of our network, we perform mean pooling along the concatenated feature map and fixed sequence length dimension. This is followed by processing through a standard dropout layer and a fully connected layer, transforming the URL features into a binary class representation for prediction. This methodology enhances the representational capability of features and improves the model's adaptability to different scale features, thereby increasing overall predictive accuracy.

\subsection{Post-training Method}
Our proposed post-training method involves further unsupervised training of LMs (specifically CharBERT in our case) using three distinct pre-training tasks: MLM, noisy LM, and DD. These tasks correspond to three distinct learning objectives: subword-level context learning, character-level context learning, and domain adaptation.

\begin{figure}
    \centering
    \includegraphics[width=0.50\textwidth]{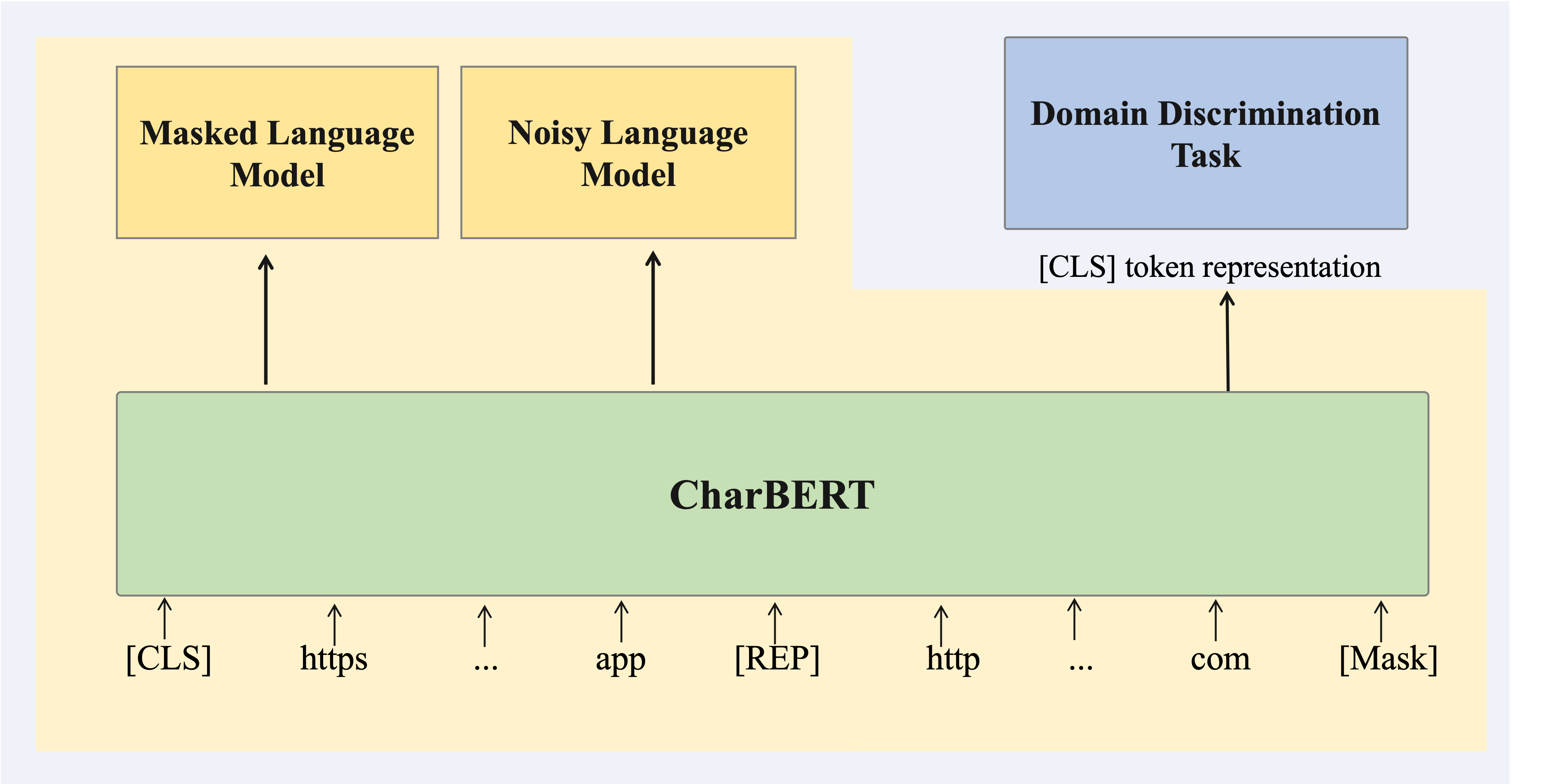}
    \caption{The diagram of three unsupervised post-training tasks. Masked LM and Noisy LM are employed for learning the contextual semantics of URL subwords and character-level representations, respectively, while the domain discrimination task is responsible for domain adaptability learning.}
    \label{}
\end{figure}

\textbf{MLM Task:}
In the post-training phase, CharBERT integrates MLM into its token-channel training, aiming to enhance its capability for contextual interpretation by inferring masked tokens within URL sequences. CharBERT also incorporates Masked Language Modeling (MLM) for its token channel training tasks. Acknowledging the sparse semantic composition of URLs, we have fine-tuned the masking ratio for MLM down from BERT's default 15\% to 10\%.

\textbf{Noisy LM Task:}
In the Noisy LM Task, our objective is to capture the internal morphological features of vocabulary terms. To this end, character noise is intentionally introduced into words, compelling the model to discern and predict the original words from their perturbed character representations.

Following prior work \cite{ma2020charbert} character noise is introduced by modifying entire words—deleting, inserting, or shuffling characters—rather than merely adjusting subwords. Consequently, this noise injection shifts the pre-training goal from subword prediction to reconstructing full, original words. To accommodate this, we constructed a new word-level vocabulary as the prediction space:

\begin{align}
H'_i &= \delta(W_6 \ast H_i + b_5); \\
p(W_j | H'_i) &= \frac{\exp(\text{linear}(H'_i)\cdot W_j)}{\sum\limits_{k=1}^{S} \exp(\text{linear}(H'_i)\cdot W_k)}
\end{align}
Here, {$\delta$ represents an activation function, $\text{linear}(\cdot)$ is a linear layer, $W_6$ is a weight matrix and $b_5$ is a bias vector, both learnable parameters of the model.} $H_i$ is the token representations from the character channel, and $S$ is the size of the word-level vocabulary.

\textbf{DD Task:}
Our proposed DD task builds upon the concept of BERT's Next Sentence Prediction (NSP) task, which traditionally determines whether two text segments follow each other in a coherent passage. The DD task adapts NSP's premise by shifting the objective towards differentiating between domains (URLs and general text domains). This requires the model to apply the vast knowledge acquired from extensive text corpora to the specific characteristics of URLs, thereby producing domain-agnostic features through adversarial training.

We format inputs as [CLS] A [SEP] B [SEP], using [CLS] and [SEP] as special tokens for initiating sequences and delimiting segments. Within this framework, half the time A and B are randomly chosen from the URL domain, tagged as TargetDomain. In the remaining cases, A is sourced from the URL domain and B from textual data, classified as MixDomain.  By employing an adversarial approach, we aim to robustly train the model toriculumnrecognize domain-specific patterns while maintaining transferability of underlying linguistic knowledge. 

Specifically, we use a domain discriminator that uses the hidden state $ h_{[CLS]} $ from the $CLS$ classification embedding to predict whether samples come from the TargetDomain or MixDomain domain. The encoder aims to map samples from both domains to a shared feature space, thereby misleading the discriminator. Specifically, before passing the hidden state $h_[CLS]$ of the classification embedding $[CLS]$ to the domain discriminator, it undergoes a gradient reversal layer (GRL). During forward propagation, the GRL behaves as an identity function. However, during backpropagation, the GRL reverses the gradient by multiplying it with a negative scalar $\lambda$. The GRL can be represented as a pseudo-function $Q_\lambda(x)$ using the following equations to describe its forward and backward behaviors:

\begin{equation}
Q_{\lambda}(x) = x 
\end{equation}

\begin{equation}
\frac{\partial Q_\lambda(x)}{\partial x} = -\lambda I.
\end{equation}

We denote the hidden state $ h_{[C L S]}$  through the GRL by  $Q_{\lambda}\left(h_{[C L S]}\right)=\hat{h}_{[C L S]} $ and then feed it to the domain discriminator as:

\begin{equation}
d=\operatorname{softmax}\left(W_{d} \hat{h}_{[C L S]}+b_{d}\right)
\end{equation}

The objective is to minimize the cross-entropy loss across all data samples originating from both the source and target domains.\vspace{6pt}
\begin{equation}
L_{d o m}=-\frac{1}{N_{s}+N_{t}} \sum_{i}^{N_{s}+N_{t}} \sum_{j}^{K} \hat{d}^{i}(j) \log d^{i}(j)
\end{equation}
Where $\hat{d}^{i} \in {0,1}$ denotes the ground-truth domain label. Through the gradient reversal layer (GRL), the parameters $\theta_{d}$ of the domain discriminator are optimized to improve its predictive capacity for domain labels.

\begin{figure*}
    \centering
    \includegraphics[width=\textwidth]{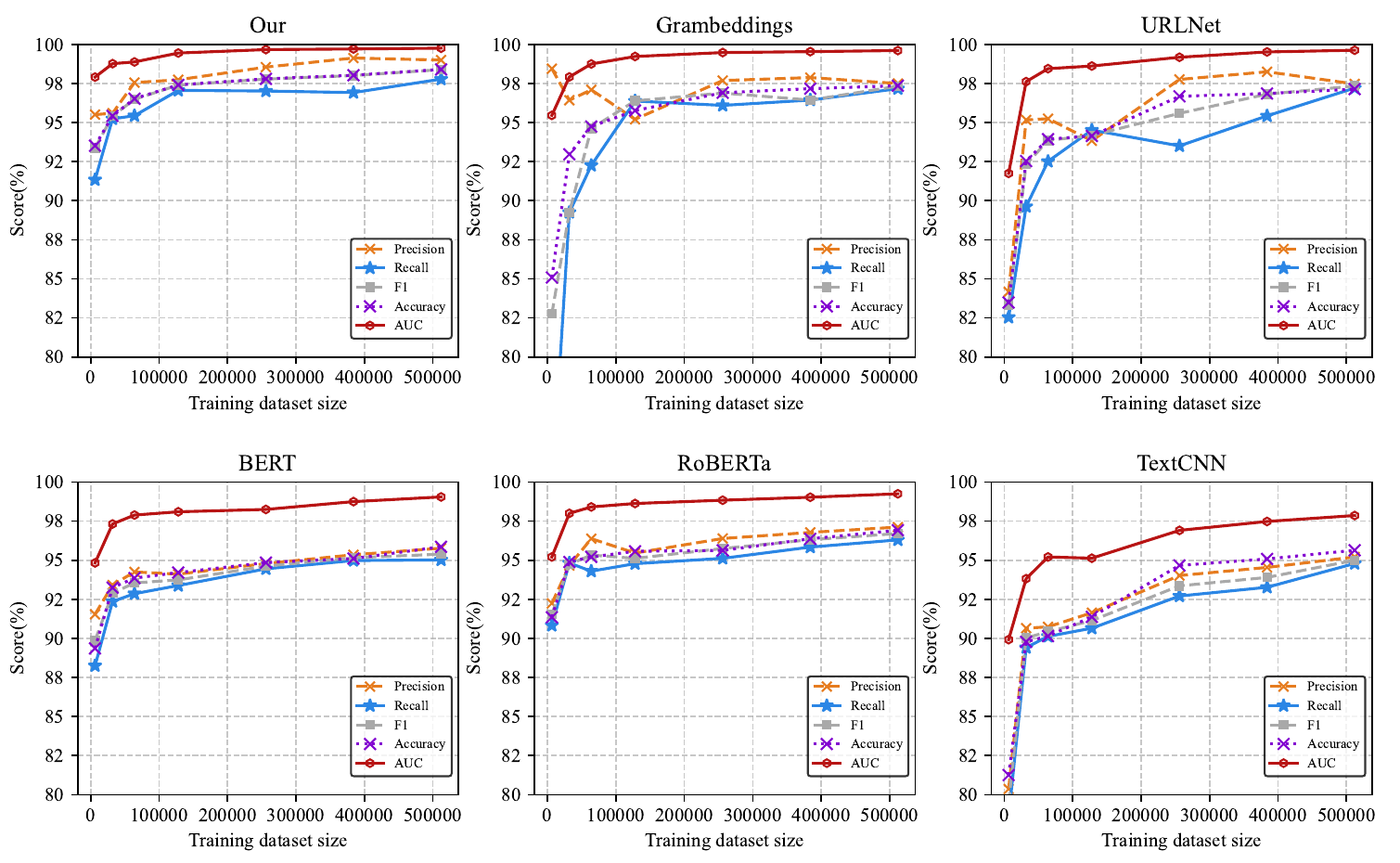}
    \caption{Performance comparison of our method against others by using the GramBeddings dataset.}
    \label{fig:fig5}
\end{figure*}

\section{Experiments}
\label{sec:experiments}
In this section, we delineate a comprehensive and structured experimental protocol to assess the efficacy of our proposed approach. The experimental configuration includes the following.

\begin{enumerate}
    \item Quantitative analysis of multi-layer feature extraction to assess feature efficacy.
    \item Experiments with varying training dataset sizes to evaluate the model's data scale dependency.
    \item Cross-dataset assessments for gauging the generalization capabilities of the model.
    \item Multi-class classification performance evaluation.
    \item Adversarial sample testing aimed at assessing model robustness.
    \item Case studies providing practical insights and detailed instance analyses.
\end{enumerate}

This comprehensive experimental design ensures a thorough and fair evaluation of the model's capabilities across various dimensions and scenarios.

 \textbf{Setup.} In our experimental setup, the backbone network, CharBERT, is an extension of BERT, pretrained on the English Wikipedia dataset (12G, 2.5 billion words). The parameters for the other three feature modules have been previously detailed. During fine-tuning, we experimented with various hyperparameters effective for downstream tasks. We ultimately selected a batch size of 64, AdamW optimizer with an initial learning rate of 2e-5 and a weight decay of 1e-4. The dropout rate was set at 0.1, with a training duration of 5 epochs. All training and inference experiments were conducted using PyTorch version 2.0, NVIDIA CUDA 11.8, and Python 3.8, with model training performed on NVIDIA A100 GPUs. For each experiment, the final model was chosen based on the best validation loss.

\textbf{Baselines.} For our experimental comparison,  we selected several state-of-the-art models, including CNN-based URLNet, TextCNN\cite{guo2019improving}, Grambeddings, and Transformer-based BERT and RoBERTa\cite{liu2019roberta}, as benchmarks against our method. To ensure fairness and reproducibility in the comparison, we retrieved their code from GitHub repositories, For binary classification tasks, we used the code without any modifications to structure or hyperparameters, applying it directly to our dataset. For multi-class tasks, we modified only the final output layer to suit specific requirements. Specifically, we chose the most complex mode, Embedding Mode 5, from URLNet for our experiments, as it demonstrated superior performance in their study.  

\subsection{Evaluation of Multi-Layer Feature}
In our approach, the multi-layer feature extraction module is designed to distill semantic features from diverse layers of the Transformer encoders in our backbone network. The core of this process lies in stacking the embedding outputs from different layers to create a multi-layered feature box. The primary objective of this experiment is to empirically validate the effectiveness of multi-layered features, demonstrating their impact on enhancing the overall model performance. This setup allows us to explore the depth and complexity of feature representations derived from different layers of the network.

In this experiment, we use the GramBeddings dataset, training our model with a random selection of 128,000 URLs, allocating 32,000 for validation, and reserving another 32,000 for testing. The results on the test set are presented in Table \ref{tab:tab3}. We assess the impact of extracting embeddings from varying depths, specifically the last 2, 3, 4, 5, and all 12 layers, on model performance. As Table \ref{tab:tab3} indicates, there is an incremental improvement in results with an increased number of layers utilized. The highest performance level is achieved when embeddings from all 12 layers are employed, underscoring the effectiveness of deeper layer integration in our model's architecture.

\subsection{Comparison with Baseline Methods}
In this section, we compare the performance of our method and baseline models in both binary and multi-class classification tasks for URL detection.

\begin{table*}
    \centering
    \caption{Performance comparison of our method against others by using the Mendeley dataset.}
    \begin{tabular*}{0.85\textwidth}{@{\extracolsep{\fill}}l@{\hspace{0.85cm}}l@{\hspace{0.8cm}}c@{\hspace{0.6cm}}c@{\hspace{0.6cm}}c@{\hspace{0.6cm}}c@{\hspace{0.6cm}}c}
    \toprule
         Training Size & Method & Accuracy & Precision & Recall & F1-score & AUC \\
    \midrule
         \multirow{6}{*}{629,184(60\%)}& BERT &0.9834  &0.8832  &0.5571 &0.6832  &0.9025 \\
         & RoBERTa &0.9865  &0.8856  &0.5638 &0.6890  &0.9142 \\
         & TextCNN &0.9723  &0.9241  &0.3578 &0.5159  &0.8756 \\
         & URLNet &0.9858  &0.9475  &0.3889  &0.5515  &0.9046 \\
         & GramBeddings &0.9801  &0.6137  &0.3026  &0.4053  &0.8205 \\
         & \textbf{Our} &0.9891  &0.8944  &\textbf{0.5786}  &\textbf{0.7027}  &\textbf{0.9209} \\
    \midrule
         \multirow{6}{*}{419,456(40\%)}& BERT &0.9821  &0.8532  &0.4803 &0.6146  &0.9220 \\
         & RoBERTa &0.9850  &0.8610  &0.5006 &0.6331  &0.9368 \\
         & TextCNN &0.9805 &0.9721  &0.2854 &0.4413  &0.8852 \\
         & URLNet &0.9842  &0.9810  &0.3019 &0.4617 &0.8992\\
         & GramBeddings  &0.9794  &0.9984  &0.0817 &0.1510 &0.8750 \\
         & \textbf{Our} &0.9878  &0.8737  &\textbf{0.5216}  &\textbf{0.6532}  &\textbf{0.9506} \\
    \midrule   
         \multirow{6}{*}{209,064(20\%)}& BERT &0.9724  &0.7715  &0.5439 &0.6380  &0.8857 \\
         & RoBERTa &0.9812  &0.8074  &0.5582 &0.6600  &0.9105 \\
         & TextCNN &0.9703 &0.9589  &0.0820 &0.1511  &0.7548 \\
         &URLNet &0.9785  &0.9653  &0.0450  &0.0860 &0.7762 \\
         &GramBeddings  &0.9804  &0.9677  &0.1306  &0.2301  &0.7869 \\
         & \textbf{Our} &0.9877  &0.8183  &\textbf{0.5743}  &\textbf{0.6749}  &\textbf{0.9187} \\
    \midrule   
         \multirow{6}{*}{104,832(10\%)} & BERT &0.9790  &0.5796  &0.5022 &0.5381  &0.8342 \\
         & RoBERTa &0.9801  &0.5958  &0.5100 &0.5496  &0.8492 \\
         & TextCNN &0.9687 &0.8425  &0.0300 &0.0579  &0.7208\\
         &URLNet &0.9789  &0.8382  &0.0735  &0.1351 &0.7584 \\
         &GramBeddings  &0.9757  &0.3879  &0.1423  &0.2082  &0.8153\\
         & \textbf{Our} &0.9822 &\textbf{0.6148} &\textbf{0.5350}  &\textbf{0.5721}  &\textbf{0.8607} \\
    \midrule   
         \multirow{6}{*}{104,832(5\%)} & BERT &0.9785  &0.5742  &0.4801 &0.5230  &0.8006 \\
         & RoBERTa &0.9798  &0.5821  &0.4885 &0.5312  &0.8139 \\
         & TextCNN &0.9547 &0.0000 &0.0000 &0.0000  &0.6230\\
         &URLNet &0.9776  &0.0000  &0.0000  &0.0000 &0.6746 \\
         &GramBeddings  &0.9782  &0.5647  &0.1139  &0.1896  &0.7752\\
         &\textbf{Our} &0.9821 &\textbf{0.6228} &\textbf{0.4932}  &\textbf{0.5505}  &\textbf{0.8101} \\
    \midrule   
         \multirow{6}{*}{10,432(1\%)} & BERT &0.9765  &0.5247  &0.3042 &0.3851  &0.7052 \\
         & RoBERTa &0.9783  &0.5362  &0.3159 &0.3976  &0.7145 \\
         & TextCNN &0.9547 &0.0000 &0.0000 &0.0000  &0.3042\\
         &URLNet &0.9776  &0.0000  &0.0000  &0.0000 &0.4419 \\
         &GramBeddings  &0.9722  &0.1808  &0.0682  &0.0991  &0.6185\\
         & \textbf{Our} &0.9812&\textbf{0.6507} &\textbf{0.3390}  &\textbf{0.4458}  &\textbf{0.7246} \\
    \bottomrule
    \end{tabular*}
    \label{tab:tab4}
\end{table*}

\subsubsection{Binary classification}
We first assess the performance of our proposed model compared to baseline methods in binary classification scenarios, using two public datasets: the balanced Grambeddings dataset and the imbalanced Mendeley dataset. To evaluate the model's dependence on training data scale, we vary the dataset sizes. In our experiments with the Grambeddings dataset, we train the model using varying training set sizes to evaluate its scalability and performance. Specifically, the training sizes we experiment with include 1\% (6,400 samples), 5\% (32,000 samples), 10\% (64,000 samples), 20\% (128,000 samples), 40\% (256,000 samples), 60\% (384,000 samples), and 80\% (512,000 samples).  For the Mendeley dataset, we train the model using varying proportions of the dataset, specifically 1\%, 5\%, 10\%, 20\%, 40\%, 60\%, and 80\%, to understand the model's performance across different training set scales.

\textbf{Results on Grambeddings dataset:} The results from the Grambeddings dataset, as illustrated in Figure \ref{fig:fig5}, consistently show that our method outperforms the baseline across all training set proportions. Notably, even with limited training data, our model exhibits remarkable proficiency.

Interestingly, with the smallest training subset of only 6,400 URLs (1\%), our method demonstrates exceptional performance, achieving an accuracy of 93.52\%, significantly higher than the baseline range of 81.25\% to 91.35\%. Moreover, our model exhibits a high recall of 91.34\%, indicating its sensitivity in identifying malicious website URLs, compared to the highest baseline recall of 90.83\%. The F1 score, a composite metric of precision and recall, reveals our method's significant advantage at 93.38\%, while the baseline methods yield F1 scores ranging from 79.47\% to 91.53\%.

Even as the baseline model's performance gradually improves with larger training samples, narrowing the gap with our method, our approach still achieves an accuracy of 98.40\% and an F1 score of 98.39\% using 80\% of the training dataset. This surpasses the best-performing baseline model, which records an accuracy of 97.36\% and an F1 score of 97.33\%. Overall, these results on the balanced dataset underscore our method's ability to maintain high accuracy and recall, effectively distinguishing between malicious and benign samples.

\textbf{Results on Mendeley dataset:} In real-world internet scenarios, legitimate web pages significantly outnumber phishing sites. Therefore, testing models on imbalanced datasets is essential. We utilize the Mendeley dataset for further evaluation. In this dataset, benign samples greatly exceed malicious ones, at a ratio of about 43 to 1, amounting to 1,561,934 URLs. This extreme imbalance presents a notable challenge to model performance, potentially leading to a bias towards numerous benign URL samples and an increased false positive rate when identifying malicious samples.

Table \ref{tab:tab4} demonstrates our method's superior performance across several metrics in a class-imbalanced scenario.  Even with just 1\% of training data, our model achieves a 44.58\% F1 score, quadrupling the baseline's performance. As training data increases, our F1 score rises to 70.27\%, vastly exceeding the baseline peak of 1.37\%. Notably, our method's superiority extends beyond F1 scores. For example, in Precision and Recall, it consistently outperforms competitors, indicating higher accuracy and lower false negatives. The AUC value of 95.06\% underlines our model's robustness in imbalanced datasets, a clear advantage over other methods. Additionally, as shown in other metrics like Accuracy and Recall, our method maintains a substantial lead, reinforcing its effectiveness in distinguishing between malicious and benign samples.

It's important to note that while URLNet and TextCNN achieve high precision on larger datasets, they suffer from a notable false positive rate, especially with smaller datasets. Grambeddings also faces difficulties, likely struggling to process rare malicious URL indicators effectively. The performance of both BERT and RoBERTa is also slightly inferior to that of our model. Our method's consistent and reliable performance across diverse data distributions signifies its effectiveness in distinguishing between malicious and benign samples, outperforming other methods in handling the challenges of imbalanced datasets.

\subsubsection{Multiple classification}
\begin{figure*}
    \centering
    \includegraphics[width=0.90\textwidth]{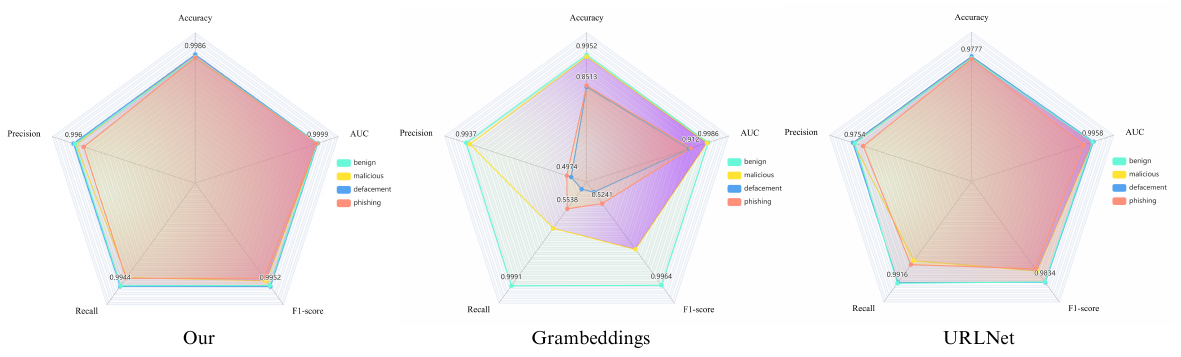}
    \captionsetup{justification=centering}
    \caption{Performance comparison of our method against others on multiple classification dataset.}
    \label{fig:fig6}
\end{figure*}

Given the evolving complexity of cyber threats, we extend our evaluation to multi-class classification, using a Kaggle 2 dataset \cite{kaggle} with four URL categories: benign (428,079), defacement (95,306), phishing (94,086), and malicious (23,645). The results, comparing our model to two baseline methods, are in Fig. \ref{fig:fig6}.

As shown in Fig.\ref{fig:fig6}, with five evaluation metrics represented at the vertices of a pentagon, our method outperforms the baseline methods across all four categories, achieving the highest performance in each metric, including an accuracy of 98.38\%. In contrast, Grambeddings falls short in multi-class scenarios, particularly in identifying defacement and phishing URLs, with an F1 score of around 50\% and an overall accuracy of 83.91\%. URLNet, while attaining an overall accuracy of 97.07\%, is also surpassed by our approach. These findings demonstrate the robustness and effectiveness of our method in complex multi-class classification tasks, marking it as a promising solution for malicious URL detection in cybersecurity.

\subsection{Cross-Dataset Testing}
Cross-dataset testing is critical for evaluating a model's adaptability to different data sources, identifying biases and limitations in the training data, and maintaining effectiveness against evolving cyber threats. We chose the best-performing model from our binary classification experiments on the Grambeddings dataset and assessed it on the Kaggle 1 dataset. The results, as shown in Table \ref{tab:tab5}, demonstrate that BERT and RoBERTa’s performances were the closest to our model, with the highest AUC reaching 96.62\%. Grambeddings, however, exhibited weaker performance, achieving lower accuracy and AUC, coupled with a high false positive rate (FPR) of 86.67\%.

Our model, in contrast, demonstrates outstanding performance across all metrics, including high precision, recall, an F1 score near 91\%, and a high AUC. Notably, it excels in FPR, at 4.59\%, outperforming all baselines. This is crucial in phishing URL detection models, emphasizing the need for low FPR to minimize false alarms in security services, ensuring benign URLs are not mistakenly classified as phishing sites.

\begin{table}
    \centering
    \caption{Cross-dataset performance generalization}
    \begin{tabular*}{\linewidth}{@{\extracolsep{\fill}}c@{\hspace{0.5cm}}l@{\hspace{0.5cm}}c@{\hspace{0.5cm}}c@{\hspace{0.5cm}}c@{\hspace{0.5cm}}c@{\hspace{0.5cm}}c}
       \toprule
         Cross-dataset & Method & Accuracy & Precision & Recall & F1-score & AUC \\
    \midrule
        \multirow{6}{*}{Gram/Kaggle} & BERT  & 0.8952 & 0.9127  & 0.8524  & 0.8815  & 0.9589 \\
        & RoBERTa  & 0.9025 & 0.9256  & 0.8684  & 0.8961  & 0.9662 \\
        & TextCNN  & 0.8702 & 0.8825  & 0.8545  & 0.8683  & 0.9166 \\
        & URLNet  & 0.8823  & 0.8947  & 0.8666  & 0.8804  & 0.9492 \\
         & Grambeddings  & 0.5214  & 0.5120  & 0.8595  & 0.6552  & 0.4647 \\
         & \textbf{Our}  & \textbf{0.9121}  & \textbf{0.9498}  & \textbf{0.8701}  & \textbf{0.9082}  & \textbf{0.9711} \\
    \bottomrule
    \end{tabular*}
    \label{tab:tab5}
\end{table}

\subsection{Adversarial Evaluation}
\label{sec:eval}
Phishing attacks, which target short-lived domains and URLs that closely resemble legitimate ones, pose significant threats. To assess our model's resilience against adversarial attacks, we employed the Compound Attack technique, as introduced by Grambeddings \cite{bozkir2023grambeddings}. This approach involves creating a real-world-compatible malicious URL from a legitimate sample by inserting evasion characters between subword tokens within the domain name. In line with the recommendations of Maneriker \etal \cite{maneriker2021urltran}, the hyphen character has been selected as the evasive token. These generated domains do not inherently exist within the pre-existing training and testing datasets but are derived from frequently observed phishing attack patterns. The process of generating adversarial samples involves using XLM-RoBERTa \cite{conneau2019unsupervised} to tag domains in given URLs, ensuring minimum tag count, randomly inserting hyphens in split parts, and swapping benign domains with malicious ones to create an adversarial list.

\begin{figure}
    \centering
    \includegraphics[width=0.40\textwidth]{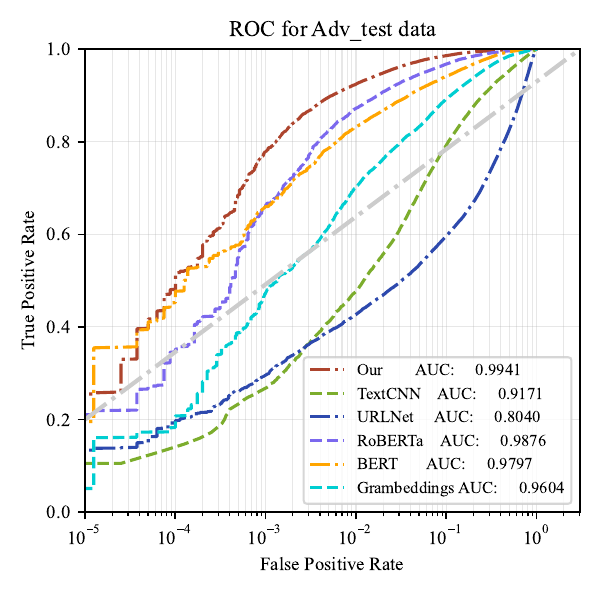} 
    \caption{ROC curve for trained models under adversarial attack.}
    \label{fig:fig7}
\end{figure}

Notably, our method involves generating a separate test dataset from the original test data, comprising 80,000 benign URLs and 40,000 malicious samples from the original set, plus an additional 40,000 newly generated adversarial malicious samples. This challenging experiment involves transforming legitimate samples into phishing URLs by hyphen insertion and tagging. We evaluate our top-performing baseline model using this new adversarial test dataset. Fig. \ref{fig:fig7} shows the ROC curves of our method and baseline methods under adversarial attacks.

Under adversarial conditions, baseline methods show significant performance drops, while our model remains stable, with an AUC of 99.41\%, about 20 percentage points higher than URLNet. Even with a fixed FPR of 0.1\%, our method achieves close to 80\% TPR, significantly surpassing the baselines, which all fall below 70\% TPR. It is observable that Transformer-based models demonstrate greater stability under adversarial attacks compared to CNN-based models, with the latter showing TPRs below 50\% at an FPR of 0.1\%. These results demonstrate our method's robustness in practice, minimizing false positives in security services even under adversarial sample attacks.

\subsection{Case Study}
\label{sec:case}
To further evaluate our model in real-world scenarios, we selected 20 active phishing URLs reported by PhishTank in October 2023, with some listed in Table \ref{tab:tab6}. Both our model and the baseline models undergo testing with those trained on 30\% of the Grambeddings dataset. Our model identified all malicious URLs with 100\% accuracy. In contrast, Grambeddings incorrectly classified 5 URLs, achieving an accuracy of 75\%, while URLNet and TextCNN each misclassified 1 URL, resulting in 95\% accuracy. BERT and RoBERTa also achieved 100\% accuracy on these samples. The Grambeddings method, despite its complex neural network combination (including CNNs, LSTM, and attention layers), does not significantly benefit in real-world scenarios, showing a considerable gap compared to prior experiments. The sensitivity of Grambeddings to minor fluctuations primarily stems from its reliance on n-gram encoding. Additionally, n-grams inherently break text into fixed-size chunks, leading to a wider, more uniform distribution of noise in the feature space. The Grambeddings method lacks specific feature refinement mechanisms, in contrast to URLNet, which employs max pooling within its CNN structure. Max pooling, though a simple operation, is instrumental in filtering out noise, allowing the model to concentrate on salient aspects of the data. 

Compared to baseline methods, our proposed approach incorporates layer-aware attention and spatial pyramid pooling. This allows for learning feature importance and filtering noise across every dimension of the input representation space. As a result, our method demonstrates enhanced accuracy in practical scenarios, particularly in identifying challenging, hard-to-distinguish, or deceptive samples. This capability marks a significant improvement in handling complex classification tasks in real-world applications.

\section{Discussion}
We briefly delineate our approach by several key aspects:

\begin{enumerate}
    \item \textbf{Generalization and Robustness}: Our method outshines existing models in a range of challenging scenarios, including small training datasets, imbalanced data, multi-classification, cross-dataset validation, and resilience to adversarial attacks, proving superior in both generalization and robustness.

    \item \textbf{Feature Engineering Simplification}: By employing a tokenizer from a pretrained language model, our approach effectively generates numerical embeddings from URL strings, streamlining the feature engineering process and enhancing semantic understanding.

    \item \textbf{Dual-channel Architectural Advantages}: Our character-aware model resolves the limitations of previous dual-path networks, like URLNet, through an interactive dual-channel architecture, facilitating better feature interaction and representation.

    \item \textbf{Efficient Training and Deployment}: Remarkably, our model achieves optimal performance within just 5 training epochs, a significant reduction compared to the 20-30 epochs typically required for baseline methods. This efficiency is attributed to the effective transfer of knowledge from the pretrained model to URL contexts.

    \item \textbf{Resource Optimization}: Despite leveraging a pretrained model, our method avoids the need for extensive retraining, thereby enabling operation on a single GPU with substantial memory capacity (over 20G). This aspect makes our model both powerful and accessible, requiring minimal computational resources for deployment.
    \item{\textbf{Theoretical Contribution:} Our study makes several significant theoretical contributions to the field of malicious URL detection:
    \begin{itemize}
        \item Our domain-specific post-training method extends the boundaries of domain-adaptive transfer learning theory, successfully adapting pre-trained language models to specialized fields. This approach provides a framework that bridges general language understanding with highly specialized tasks such as URL analysis.
        \item Our research advances the theoretical understanding of how to effectively represent and integrate information at various levels of abstraction within deep Transformer models. This contribution is particularly relevant to complex, layered architectures like BERT when applied to structured data such as URLs.
        \item We contribute to the theoretical understanding of how the importance of features can be dynamically assessed and utilized in deep learning models. Our layer-wise attention mechanism offers new insights into the relative contributions of different abstraction levels in complex neural network structures.
    \end{itemize}}
    \item {\textbf{Future Direction:} Our research has demonstrated the benefits of integrating pre-trained models with multi-scale (local and global) and multi-dimensional (character-level and subword-level) information for URL analysis. While the potential for further information extraction from URLs themselves may be approaching its limits, there remains significant unexplored territory in the analysis of the web pages these URLs point to. Future research should focus on extracting, representing, and fusing information from various webpage elements, such as: webpage content, visual elements (e.g., website icons, page layouts, and design patterns), HTML structure and JavaScript code. This holistic approach has the potential to significantly improve our ability to identify and mitigate evolving cyber threats by leveraging the rich, multi-faceted nature of web content. } 

\end{enumerate}

\begin{table}
    \centering
    \caption{Example of malicious URLs from \textit{PhishTank}}
    \begin{tabularx}{0.9\linewidth}{X}
    \toprule
        \textbf{Malicious URL}\\
    \midrule
        https://bersw6.wixsite.com/my-site\\
        https://www.lloyds.user-review24.com/\\
        http://798406.selcdn.ru/webmailprimeonline/index.html\\
        https://bansruasdras.web.app/\\
        https://cloudflare-ipfs.com/ipfs/bafybeibfyqcvrjmwlpip\\
        qkdyt2xr46cea7ldciglcbybfwtk7cieugcj3e/\\
        https://cion3.net/-/pdf/js/\\
    \bottomrule
    \end{tabularx}
    \label{tab:tab6}
\end{table}

\section{Conclusion}
\label{sec:conclusion}
In this paper, we present a high-performance method for detecting malicious URLs, leveraging the advantages of pretrained LM and advanced Transformer networks, augmented with three closely integrated feature learning modules for enhanced URL feature extraction. Key benefits of our approach include: 1)End-to-end learning capability, requiring only raw URL strings without manual preprocessing; 2)An interactive subword and character-level feature learning network architecture for improved character-aware subword representations; 3) Effective multi-level and multi-scale URL feature learning based on our proposed feature learning modules. Our method, extensively evaluated on various URL datasets, consistently outperforms existing state-of-the-art baseline methods, producing stable decisions across scenarios. Notably, it demonstrates superior generalization and robustness in cross-dataset detection and adversarial sample attacks, enhancing its reliability in practical applications. Additionally, in case study, our method accurately identifies all active malicious webpages, further evidencing its efficacy.

\printcredits

\bibliographystyle{splncs04}

\bibliography{mainv2}

\end{document}